\documentclass[twocolumn,superscriptaddress,amsmath,amssymb,showpacs,floatfix,preprintnumbers]{revtex4-1}

\usepackage{amsmath}
\usepackage{amssymb}
\usepackage{graphicx}
\usepackage{color}
\usepackage[colorlinks,citecolor=blue,linkcolor=blue]{hyperref}
\usepackage{epsfig}
\usepackage{amsmath}
\usepackage{amssymb}
\usepackage{epstopdf}
\usepackage{sidecap}
\usepackage{booktabs}
\usepackage{rotating}
\usepackage[bottom]{footmisc}

\sidecaptionvpos{figure}{c}

\DeclareGraphicsExtensions{.png .jpg .pdf}

\newcommand{\s}{\sigma}
\newcommand{\mc}{\mathcal}
\newcommand{\mb}{\mathbf}
\newcommand{\mr}{\mathrm}

\newcommand{\dql}{\textquotedblleft}
\newcommand{\dqr}{\textquotedblright}

\begin{document}

\title{Magnetism in twisted triangular bilayer graphene quantum dots}

\author{M. Mirzakhani}\email{mirzakhani@ibs.re.kr}
\affiliation{Center for Theoretical Physics of Complex Systems, 
Institute for Basic Science, Daejeon, 34126, South Korea}

\author{H. C. Park}\email{hcpark@ibs.re.kr}
\affiliation{Center for Theoretical Physics of Complex Systems, 
Institute for Basic Science, Daejeon, 34126, South Korea}
\author{F. M. Peeters}\email{francois.peeters@uantwerpen.be}
\affiliation{Department of Physics, University of Antwerp, 
Groenenborgerlaan 171, B-2020 Antwerp, Belgium}
\affiliation{Departamento de Fisica, Universidade Federal do
	Cear\'a, Campus do Pici, 60455-900 Fortaleza, Cear\'a, Brazil}
\author{D. R. da Costa}\email{diego\_rabelo@fisica.ufc.br}
\affiliation{Departamento de Fisica, Universidade Federal do
	Cear\'a, Campus do Pici, 60455-900 Fortaleza, Cear\'a, Brazil}


\begin{abstract}

Using a tight-binding model along with the mean-field Hubbard method,
we investigate the effect of twisting angle on the 
magnetic properties of twisted  bilayer graphene (tBLG) quantum dots (QDs) with  
triangular shape and zigzag edges.
We consider such QDs in two configurations: when their initial untwisted structure 
is a perfect AA- or AB-stacked BLG, referred to as AA- or AB-like dots.
We find that AA-like dots exhibit an antiferromagnetic spin polarization for small twist
angles, which transits to a ferromagnetic spin polarization beyond a critical twisting 
angle $\theta_c$.
Our analysis shows that $\theta_c$ decreases as the dot size increases, 
obeying a criterion, according to which once the maximum energy
difference between electron and hole edge states (in the single-particle picture) is less than 
$(U / \gamma_0)\, t_0$, the spin-polarized energy levels are aligned ferromagnetically
[$U$ is the Hubbard parameter and $\gamma_0$ ($t_0$) the graphene intralayer (interlayer) hopping].
Unlike AA-like dots, AB-like dots exhibit finite magnetization for any twist angle.
Furthermore, in the ferromagnetic polarization state, the ground net spin for both dot
configurations agrees with prediction from Lieb's theorem.

\end{abstract}

\maketitle


\section{Introduction} \label{intro}

Magnetic materials are critical components for a wide range of technological applications.
Due to the outstanding electronic and structural properties of graphene 
\cite{Geim2007,Castro2009,Vozmediano2010}, it has 
also attracted huge amounts of research attention to the magnetism associated with 
carbon-based materials since its first isolation in 2004 \cite{Novoselov2004}.
While ideal graphene itself does not show magnetic properties, several of its derivative
materials and nanostructures, both realized in practice and studied in theory, exhibit 
various forms of magnetism 
(see, e.g., Refs.~\cite{Nakada1996,Son2006,Fernandez2007,Yazyev2010,Potasz2012,Oteyza2022}).
For example, both theoretical \cite{Duplock2004,Palacios2008} and experimental
\cite{Ugeda2010,Gonzalez2016} studies reveal that a defective graphene
with some $p_z$ electrons missing from its crystallographic lattice displays a net spin.
Theoretical studies, on the other hand, predict that a wide range of finite nanostructured 
graphene exhibit magnetic ordering.
Triangular graphene quantum dots (QDs) [the corresponding PAH (polycyclic aromatic hydrocarbon)
molecule is known as 
[n]-triangulenes, where $n$ is the number of hexagons along each molecular edge] and 
graphene nanoribbons with zigzag edges are iconic examples of such structures
\cite{Nakada1996,Son2006,Yazyev2010,Oteyza2022}.
The origin of magnetism in nanostructured graphene (generally in carbon-based structures) 
as a light material is mainly related 
to the imbalance of sublattice atoms \cite{Fernandez2007,Yazyev2010}, which is 
different from other conventional magnetic materials like Ni, Fe, and Co.
Because of this property, graphene $\pi$ magnetism is more delocalized and isotropic than
conventional magnetic.

The huge progress achieved within the last few years in the fabrication of graphene 
nanostructures has provided unprecedented opportunities for the synthesis and 
characterization of such type of materials.
Many carbon-based nanostructures, such as triangulene \cite{Pavlicek2017,Mishra2019,Su2019}, 
zigzag-edged graphene nanoribbons \cite{Tao2011,Ruffieux2016}, and Clar's goblet \cite{Mishra2020},
whose intrinsic magnetic properties were theoretically predicted previously
\cite{Fernandez2007,Yazyev2010,Nakada1996}, have been synthesized and studied over the last 
few years.
Despite the lack of conclusive evidence in the first two mentioned graphene nanostructures, 
magnetism in Clar's goblet was recently demonstrated in Ref.~\cite{Mishra2020}.
Ferromagnetism in twisted bilayer graphene (tBLG) has also been recently reported
\cite{Sharpe2019}, which demonstrates the fascinating advances reached in 
carbon-based magnetism.
Further details of recent experimental progress in nanostructured graphene materials
that either display or have the potential to trigger its magnetic properties
(mostly zigzag-edged graphene flakes) 
can be found in recent review-articles \cite{Oteyza2022,Liu2020,Song2021}.
These new advancements in the synthesis of graphene nanostructures motivated us to study the
magnetic properties of QDs in tBLG.

\begin{figure}
	\centering
	\includegraphics[width = 8. cm]{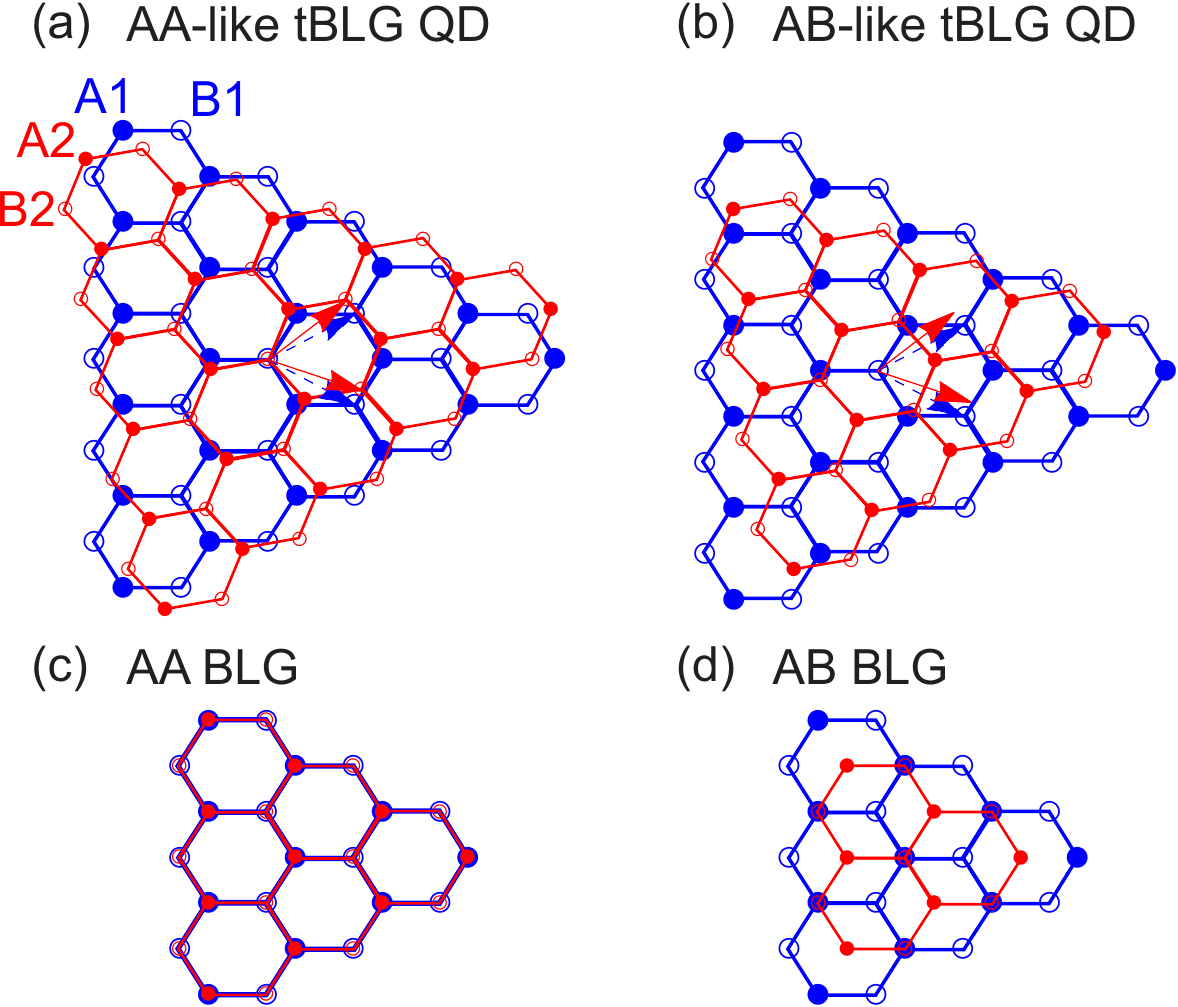}
	\caption{Schematic geometry of the triangular tBLG QDs with lateral	zigzag edges.
		The layers are depicted in two different colors: 
		blue (bottom) and red (top).
		Filled (empty) circles in each layer indicate A$i$ (B$i$) [$i=1,2$]
		sublattice.
		(a) An AA-like dot, whose untwisted arrangement
		corresponds to the AA-stacked triangular BLG dot [panel (c)], in which 
		each carbon atom of the top layer is placed exactly above the 
		corresponding atom of the bottom layer.
		(b) An AB-like dot, whose untwisted arrangement corresponds to the AB-stacked
		triangular BLG dot [panel (d)], in which different sublattices of two 
		layers are directly
		coupled to each other (A1-B2) so that the two others are situated above the 
		centers of the opposing layer hexagons (B1,A2).
	}
	\label{fig1}
\end{figure}

At this point, it needs to be noticed that stacking two or more layers of graphene can 
have a significant impact on its 
mechanical, electronic, and magnetic properties, both in bulk and nanostructured forms,
see, e.g., Refs. \cite{Velasco2014,Pereira2007,Zarenia2013,daCosta2015,Belouad2016,Guclu2011,
Sahu2008,Henriksen2008,Zhang2010,Velasco2014,Fang2015,Rozhkov2016re,
Mirzakhani2016,Nascimento2017}.
In the case of BLG QDs, it has been demonstrated that the edges and geometries 
play an essential role in 
modifying the energy spectrum \cite{daCosta2016a,daCosta2016b} as well as its magnetic properties
\cite{Nascimento2017}, similar to monolayer graphene (MLG) QDs 
\cite{Zhang2008,Guclu2014Re}.
During the last decade, two well-known stacks of BLG QDs, i.e., 
the AA and AB types, have been experimentally realized and extensively studied 
\cite{Allen2012,Goossens2012,Eich2018,Ge2020}.
Besides, the effect of twisting on the electronic and transport properties of BLG QDs
has also been recently addressed both theoretically
\cite{Landgraf2013,Tiutiunnyka2019,Mirzakhani2020,Han2020,Tepliakov2020,Wang2021,Wang2022} 
and experimentally \cite{Zhou2021}. 
These results, for example, suggest that the twisting axis can be utilized to tune the interlayer 
conductance \cite{Han2020} or that the twist angle can modify the energy levels \cite{Tiutiunnyka2019,Mirzakhani2020} in stacked graphene nanostructures.
Despite several theoretical studies pertinent to the electronic properties of 
tBLG QDs, to our knowledge, there is currently no theoretical study on 
the magnetic properties of such QDs. 

In this paper, we aim to investigate the effect of twisting angle on the magnetic 
properties of a triangular QD in tBLG. 
As previously demonstrated in Refs.\ \cite{Fernandez2007,Yazyev2010}, zigzag edges host low-energy
edge states, which causes magnetism to develop in carbon nanostructures.
In this study too, we only address triangular tBLG QDs with well-defined zigzag edges as shown 
in Fig.~\ref{fig1}.
Combining the tight-binding model (TBM) and the electron-electron (e-e) interactions
addressed self-consistently at the level of the mean-field (MF) Hubbard model, we 
numerically investigate how twisting angle affects the magnetic ordering in 
(triangular) tBLG QDs.
To this end, we study systematically two configurations of tBLG QDs: AA-like [Fig.~\ref{fig1}(a)] 
and AB-like QDs [Fig.~\ref{fig1}(b)],
whose untwisted arrangements correspond, respectively, to the ideal AA- and AB-stacked
BLG QDs, as depicted in Figs.~\ref{fig1}(c) and \ref{fig1}(d).

\begin{figure}
	\centering
	\includegraphics[width = 8.5 cm]{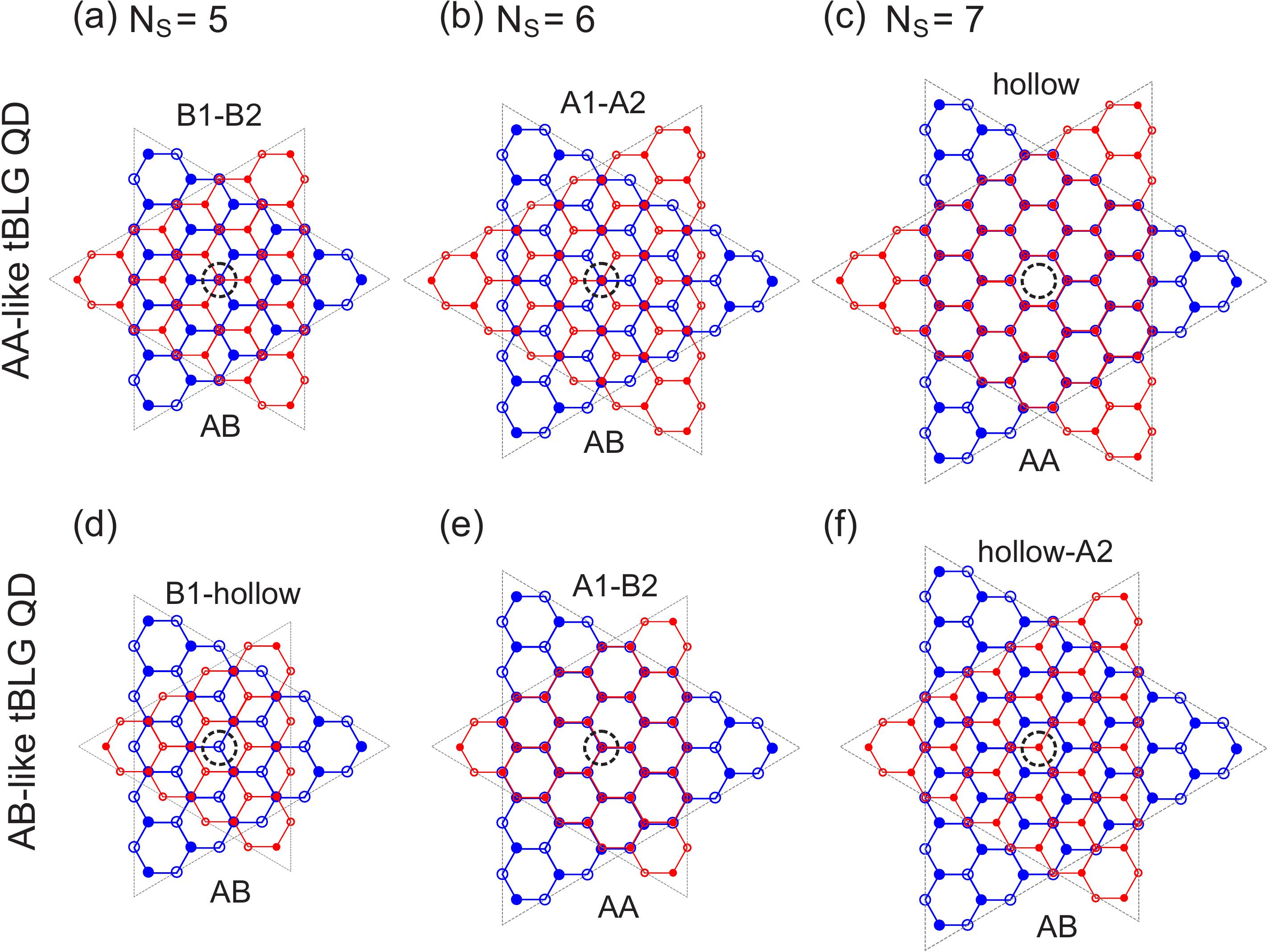}
	\caption{Geometries and geometric centers (GC, dashed circles) of 
		three different dot sizes with the edge atoms of  $N_s = 5,6,$ and $7$
		[respectively, representing $N_s^{(2),(3), (4)}$-group dots with 
		$n = 1$ in Eq.~\eqref{eq:categ}] 
		at a twisting angle of $\theta = 60^\circ$ for the AA-like (upper panels) and 
		AB-like (lower panels) dots.
		In each case, the GC (as the twisting point, indicated at the top of each panel) is
		determined by the lattice sites of the bottom and top layers where the GC is located.
		Notice that, at $\theta = 60^\circ$, both dots can result in AA or AB arrangements 
		in the layers' intersection.
		The same patterns are repeated for different dot sizes, characterized by 
		Eq.~\eqref{eq:categ}.
	}
	\label{fig2}
\end{figure}

Interestingly, our numeric calculations predict a magnetic quantum phase transition at a critical 
twisting angle for AA-like dots.
We find that the AA-like dots exhibit an antiferromagnetic phase at small twisting 
angles, which beyond a critical angle $\theta_c$ transition to a ferromagnetic phase occur
for which the total spin $S$ agrees with Lieb's theorem \cite{Lieb1989}.
Lieb's theorem predicts the total spin of the Hubbard model's ground state in bipartite
lattices.
Our analysis shows that $\theta_c$ decreases as the dot size increases.
We also find a criterion for the value of $\theta_c$, according to which once the maximum energy
difference between electron and hole edge states in the single-particle (SP) picture is less than 
$(U / \gamma_0) \times t_0$, the spin-polarized energy levels are aligned ferromagnetically.
Here, $U$ is the Hubbard parameter and $\gamma_0$ ($t_0$) denotes the graphene intralayer 
(interlayer) hopping.

Unlike AA-like dots, there is no phase transition from an antiferromagnetic to a ferromagnetic 
phase in AB-like dots, and the spins in such dots are ferromagnetically polarized all 
twist angles.
Furthermore, in the ferromagnetic phase, the net spin of the studied tBLG QD configurations scales 
linearly with dot size by one spin unit; nevertheless, AA-like dots result in an integer net spin 
and AB-like ones in a half-integer.

The paper is organized as follows: 
In Sec.\ \ref{theo}, we present the QD structures in tBLG and the basics of our 
numerical method.
Section \ref{num} is dedicated to results and discussions.
A summary and concluding remarks follow in Sec.\ \ref{con}. 

\section{Theory and model} \label{theo}
In this paper, we study the intrinsic magnetism of triangular tBLG QDs for two types of
configurations.
Figures \ref{fig1}(a) and \ref{fig1}(b) depict two possibilities for creating a tBLG QD from two 
comparable monolayer QDs.
A zigzag triangular tBLG QD [Figs.~\ref{fig1}(a)], which is built from the two 
perfectly flat triangular MLG QDs with the same shape, size, and edge boundaries in 
which the second MLG QD (top) is rotated by an angle $\theta$ around the geometric 
center of the dot.
In this case, untwisted arrangement ($\theta = 0$) corresponds to an AA-stacked BLG QD
configuration [Fig.~\ref{fig1}(c)].
We will refer to such a structure as an \dql AA-like dot\dqr.
Another configuration choice is shown in Fig.~\ref{fig1}(b),
in which the top layer is smaller (one atom at the edge) than the bottom layer, and 
$\theta = 0$ corresponds 
to a perfect AB-stacked BLG QD, see Fig.~\ref{fig1}(d).
Such a structure is referred here to as an \dql AB-like dot\dqr.
In both configurations, the interlayer spacing is $d_0 = 0.335$ nm.
Each dot can be characterized by the number of atoms on one edge of the (bottom) layer,
$N_s$. 
The total number of carbon atoms in a layer of such triangular dots with zigzag edges is 
$N_{L} = N_s^2 + 4 N_s  + 1$.
Notice that in zigzag triangular graphene QDs, all edge atoms belong to the same sublattice, 
which here is B sublattice in both layers.

Depending on the dot size, the geometric center (through which the twist 
axis passes) coincides at different positions, and each dot can result in either an 
AA- or an AB-BLG configuration when $\theta = 60^\circ$.
This feature is illustrated in Fig.~\ref{fig2}. 
Accordingly, from a geometrical standpoint, we categorize the dots into three groups, 
identifying them by the edge atom numbers as
\begin{equation} \label{eq:categ}
  N_s^{(p)} = p + 3n  \quad  (n = 0,\ 1,\ \ldots),
\end{equation}
where, $p = 2,3, \text{and } 4$.
For more details, see Fig.~\ref{fig2} and the caption therein.

In order to study the magnetic properties of tBLG QDs, we use the widely applied 
one-orbital MF Hubbard model 
\cite{Yazyev2010,Fernandez2007,Oteyza2022,Wolf2021,Phung2022,Li2022}.
This model considers only the unhybridized $p_z$ atomic orbital of the carbon atoms.
The $p_z$-electron states govern all low-energy features of graphene, both electronic 
and magnetic.
The Hubbard model can be expressed as the sum of two terms \cite{Hubbard1963},
\begin{equation}  \label{eq:Htot}
  \mc H = \mc{H}_0 + \mc{H}_\mr{U}.
\end{equation}
The first term ($H_0$) is the SP TB Hamiltonian, which in the second quantization 
formalism can be written as
\begin{equation}  \label{}
  \mc{H}_0 = \sum_{i,\s} \epsilon_{i \s} c^\dagger_{i \s} c_{i \s}
      -\sum_{\langle i,j \rangle, \s} [t(\mathbf{d}_{ij}) c^\dagger_{i \s} c_{j \s} 
      + \text{h.c.}],
\end{equation}
where $c^\dagger_{i \s}$ and $c_{i \s}$ are, respectively, the creation and annihilation 
operators for an electron on lattice site $i$ with on-site energy $\epsilon_{i \s}$
(we set $\epsilon_{i \s} = 0$).
$\mathbf{d}_{ij}=\mb{r}_i-\mb{r}_j$ is the distance between 
the lattice points ($\mb{r}_i$, $\mb{r}_j$), $t(\mb{d}_{ij})$ 
is the corresponding transfer integral, and $\langle i,j \rangle$ indicates a 
summation over nearest-neighbor sites.
Using the Slater-Koster form, the transfer integral between the atoms $i$
and $j$ can be written as \cite{Slater1954,Nakanishi2001,Uryu2004,Laiss2010,Moon2013},
\begin{equation}
  -t(\mb{d}_{ij}) = V_{pp\pi} \Big[1 - \Big(\frac{\mb{d}_{ij}.\mb{e}_z}{d_{ij}} 
  \Big)^2 \Big]
  + V_{pp\sigma} \Big( \frac{\mb{d}_{ij}.\mb{e}_z}{d_{ij}} \Big)^2.
\end{equation}
Here, $V_{pp\pi} = \gamma_0 \exp[- (d_{ij}-a_{cc}) / \delta_0]$ and 
$V_{pp\sigma} = t_0 \exp[-(d_{ij}-d_0) / \delta_0],$ where 
$a_{cc} = 0.142$ nm is the carbon-carbon distance of graphene and $\delta_0 = 0.184\,a$ 
($a = \sqrt{3}\, a_{cc}$ is the graphene lattice constant) is the decay length.
$\gamma_0 \approx -2.7$ eV and $t_0 \approx 0.48$ eV are the
intralayer and interlayer nearest-neighbor hopping parameters, respectively.
For the intralayer coupling, we include only the nearest-neighbor hopping parameter.
But for the interlayer coupling, since the layers are rotated and the neighbors are not 
on top of each other, we take the interlayer coupling terms for atomic distances of
$d_{ij} \leqslant 4a_{cc}$.
For $d_{ij} > 4a_{cc}$, the transfer integral is exponentially small and can be safely 
ignored \cite{Moon2013}.
The electron-hole symmetry is also broken as a result of the mixing between the two sublattices.

The second term in Hamiltonian \eqref{eq:Htot} is the Hubbard term that introduces 
e-e interactions through the repulsive on-site Coulomb interaction,
\begin{equation}  \label{eq:Hub}
  \mc{H}_\mr{U} = U \sum_i n_{i \uparrow} n_{i \downarrow},
\end{equation}
where $n_{i \s} = c^\dagger_{i \s} c_{i \s}$ ($\s = \uparrow, \downarrow$) is the 
spin-resolved electron density at site $i$.
The parameter $U > 0$ is the Hubbard parameter and denotes, in the short-range
regime, the on-site Coulomb repulsion energy for each pair of electrons with opposite 
spins on the same site $i$. 

\begin{figure*}
	\centering
	\includegraphics[width = 16 cm]{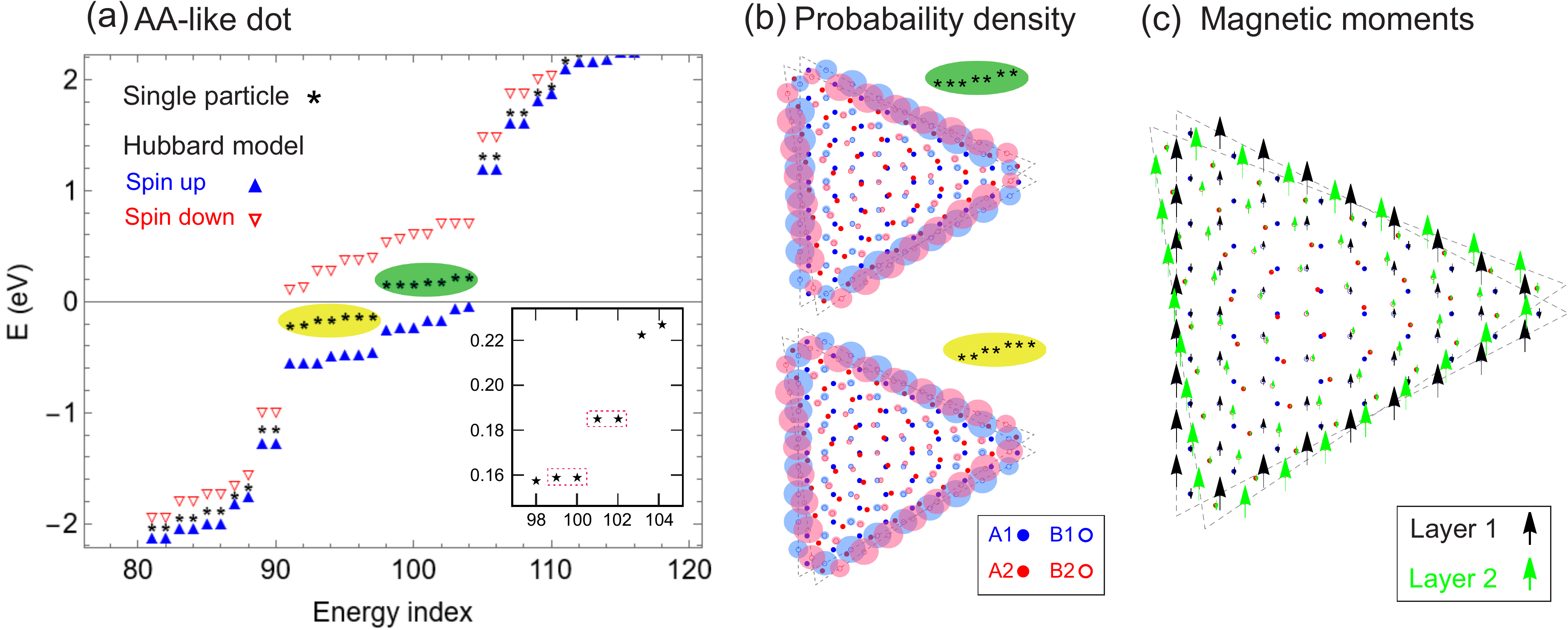}
	\caption{(a) Energy levels of an AA-like triangular tBLG QD with $N_s = 8$ and twisting 
		angle of $\theta = 7^\circ$ in the frame of 
		SP (single particle, black stars) and Hubbard model (triangles) 
		around the Fermi energy $E_f = 0$.
		Filled blue (empty red) triangles depict the spin up (down) energy levels.
		The inset shows a zoom of the LUMO cluster energy levels
		(encircled by the green oval in the main figure), indicating two doubly degenerate 
		(red boxes) and	three non-degenerate energy levels.
		(b) Probability densities corresponding to two clusters of the LUMO 
		(upper panel) and HOMO (lower panel) energy levels around $E_f$.
		(c) Local magnetic moments $m_i$ shown by the black and green arrows
		corresponding to the lower and upper layers, respectively.
		The length of the arrows signifies the relative magnitude of the magnetic
		moments.
	}
	\label{fig3}
\end{figure*}

In the MF approximation, the Hubbard term \eqref{eq:Hub} at half-filling can be
rewritten as 
\begin{multline}  \label{eq:mfHu}
  \mc{H}_\mr{U} = U \sum_i \big(n_{i \uparrow} - \frac{1}{2} \big) \big(n_{i \downarrow} 
  - \frac{1}{2} \big)\\
  = U \sum_i \big[\langle n_{i \downarrow} \rangle n_{i \uparrow} +
  \langle n_{i \uparrow} \rangle n_{i \downarrow} - 
  \frac{1}{2} (n_{i \uparrow} + n_{i \downarrow}) - \langle n_{i \uparrow} \rangle
  \langle n_{i \downarrow} \rangle + \frac{1}{4} \big].
\end{multline}
Here, a spin-up electron, $n_{i \uparrow}$, at site $i$ interacts with the average 
density of spin-down
electrons $\langle n_{i \downarrow} \rangle$ at the same site and vice versa.
Accordingly, the MF Hubbard Hamiltonian only contains SP operators. 
It is also worth mentioning that such MF approximation shows the Hartree term, 
that is,
the MF Hubbard term is only written for the $z$ component of the spin moment.
This is the most common method for examining a system's magnetic properties
\cite{Yazyev2010,Fernandez2007,Oteyza2022}.

To solve the problem for $\mc H$, we use self-consistent calculations that start 
with randomly chosen initial values for the unknown electron densities
$\langle n_{i \s} \rangle$.
The Hamiltonian \eqref{eq:Htot} is then diagonalized to obtain the new eigenvalues 
and eigenvectors, which are used to compute the new spin densities 
$\langle n_{i \uparrow} \rangle$ and $\langle n_{i \downarrow} \rangle$ on each site.
Then, the obtained new spin densities are fed as the initial values for the 
next iteration.
The procedure is repeated until all values of $\langle n_{i \s} \rangle$ are converged.
The convergency criterion is met when 
$\langle n_{i \s} \rangle^{s+1} - \langle n_{i \s} \rangle^{s} < \delta$, where 
$\delta$ is a small number chosen $\delta = 10^{-6}$ and $s$ is the 
self-consistent cycle index.
After achieving self-consistency, one can compute the magnetic moment per 
atomic site 
\begin{equation} \label{eq:mom}
  m_i = \frac{\langle n_{i \uparrow} \rangle - \langle n_{i \downarrow} \rangle}{2},
\end{equation}
and the total spin of system $S = \sum_i m_i$.

From literature (see, e.g., Refs.~\cite{Yazyev2010,Nascimento2017}), there is currently no
consensus on the exact value of $U$ in the case of graphene-based structures.
Such a parameter should ideally be approximated using experimental data, and 
there are currently no standard or direct experiments on magnetic graphene systems 
to which we may refer.
However, it has been demonstrated that for specific values of $U / |\gamma_0|$, 
the results of MF Hubbard model calculations are in good agreement with results 
from first-principles approaches based on density functional theory
\cite{Fernandez2007,Pisani2007,Gunlycke2007}.
In general, the most common range of $U$ parameter values is $U \sim 3.0 - 3.5$ eV,
which corresponds to $U / |\gamma_0 | \sim 1.1-1.3$ \cite{Yazyev2010}.†
At $U / |\gamma_0| \approx 2.23$, ideal graphene undergoes a
Mott-Hubbard transition into an antiferromagnetically ordered insulating phase
\cite{Sorella1992,Fujita1996}.
In our calculations, we use the value of $U = 1.2\, |\gamma_0 | = 3.24$ eV, unless otherwise 
specified.

\begin{figure*}
	\centering
	\includegraphics[width = 18 cm]{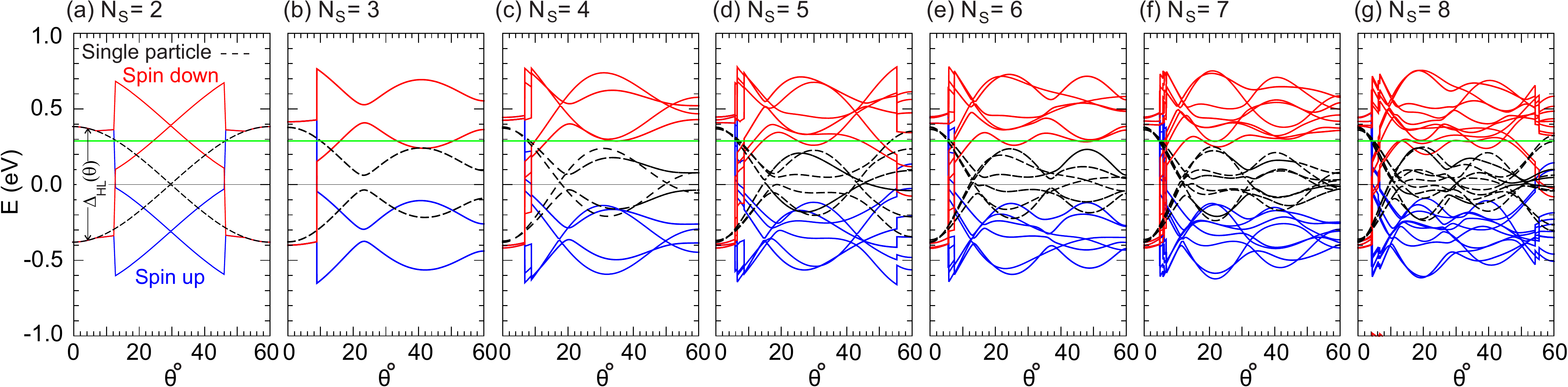}
	\caption{Energy levels of the AA-like dot as a function of the twist angle $\theta$
	for different dot sizes as indicated by the $N_s$ value in each panel. 
	The results are presented for SP (dashed black curves) and Hubbard 
	models [spin up (blue) and spin down (red)] around the Fermi energy 
	$E_f = 0$.
	The solid green lines indicate $(q \times t_0) / 2$ [$q \equiv U / |\gamma_0| = 1/2$; explained in 
	Eq.~\eqref{eq:phase}].
	As clearly seen, when $\Delta_{HL} (\theta) / 2$ falls below the green line, the 
	energy levels become spin polarized in all cases.
	}
	\label{fig4}
\end{figure*}

\begin{figure}
	\centering
	\includegraphics[width = 8.5 cm]{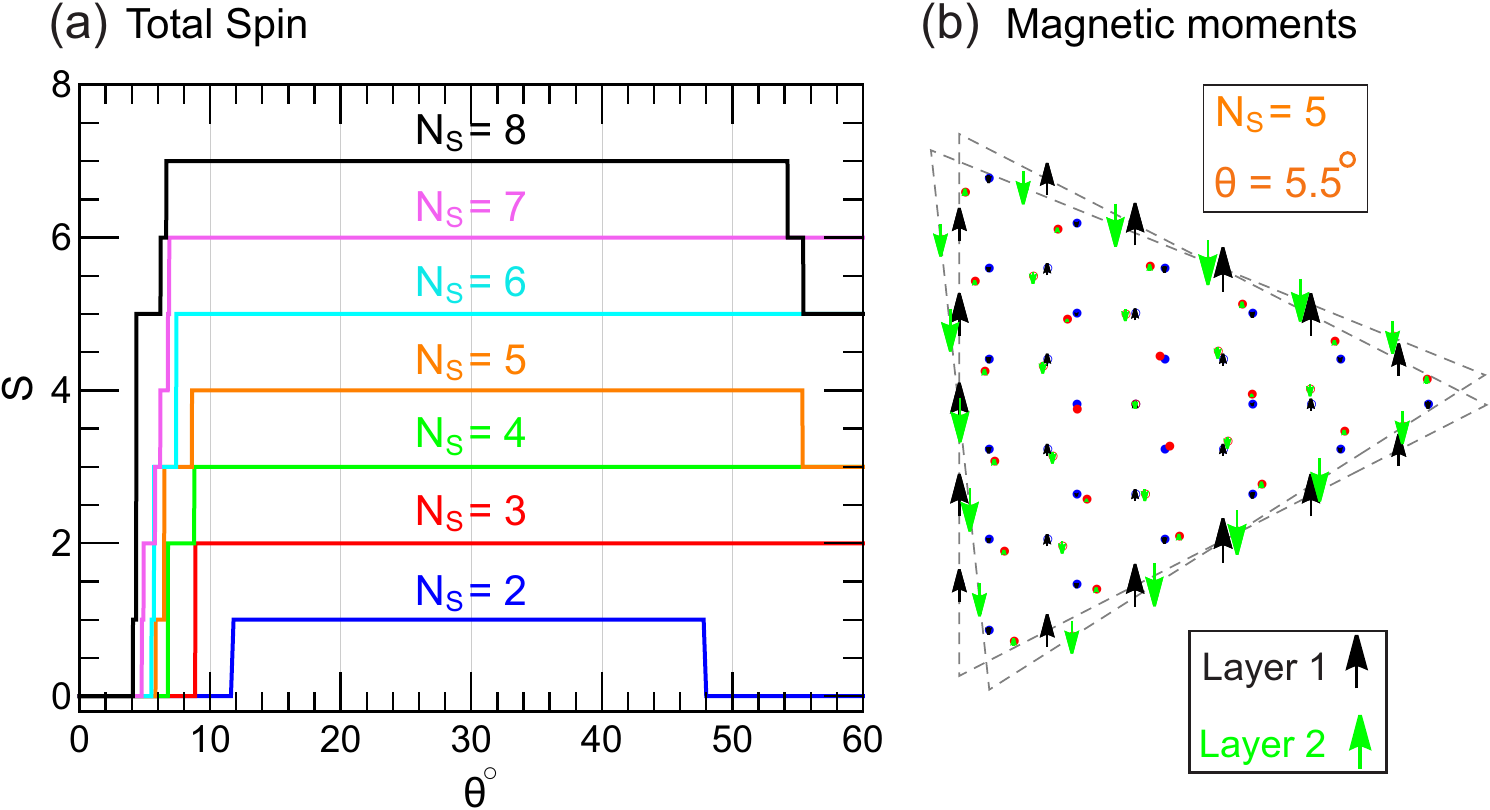}
	\caption{(a) Total spin $S$ as a function of $\theta$ for AA-like dots whose energy
	levels are shown in Fig.~\ref{fig4}.	
	For certain twisting angles, $S$ achieves a value that is 
	consistent with Lieb's theorem prediction [Eq.~\eqref{eq:Lieb}].¬†
	(b) Local magnetic moments for a dot with $N_s = 5$ and $\theta = 5.5^\circ$, presenting
	an antiferromagnetic phase with $S = 0$, as seen in panel (a).
	}
	\label{fig5}
\end{figure}

\section{Numerical results} \label{num}
\subsection{AA-like triangular tBLG QDs}

First, we consider the AA-like dots as illustrated in Fig.~\ref{fig1}(a).
Figure \ref{fig3}(a) shows the energy spectrum of an AA-like dot with $N_s = 8$ edge atoms  and 
a twist angle of $\theta = 7 ^\circ$ as a function of the energy index.
The results are presented for the SP (black stars) and the MF Hubbard models near the 
Fermi energy $E_f = 0$. 
Filled (empty) triangular symbols correspond to the spin up (down) energy states.
As seen, in the case of SP energy levels, there are two clusters of nearly-degenerate
energy levels [indicated by yellow and green ovals in Fig.~\ref{fig3}(a)]
corresponding to the highest occupied and lowest unoccupied molecular 
orbitals (HOMOs and LUMOs) around $E_f = 0$.
Such electronic states originate from sublattice imbalance of each layer of the dot.
In a bipartite structure, one can find $N_Z$ \dql strict\dqr\ zero-energy states 
(according to the \dql benzenoid graph\dqr\ theory \cite{Fajtlowicz2005}) equating 
to sublattice imbalance $|N_A - N_B|$,
where $N_A$ and $N_B$ are the number of sites in sublattices A and B, respectively
\cite{Yazyev2010,Oteyza2022}.
In the case of zigzag triangular MLG QDs, the sublattice imbalance is proportional to 
the number of atoms at one edge, i.e., $|N_A - N_B| = N_s - 1 = N_Z$ \cite{Potasz2010}.
Here, two clusters of $N_Z = 7$ nearly-degenerate energy states (totally 14 states) 
appear around $E_f$,
due to the two triangular MLG QDs, which are gaped as a result of the interlayer 
coupling between the edge atoms of the two dot layers.
The SP energy gap is $\Delta_{SP} \approx 0.28$ eV.
Notice that each cluster of energies consists of two doubly degenerate and three
non-degenerate states as shown in the inset of Fig.~\ref{fig3}(a).
Probability densities corresponding to each cluster of electron and hole energy states
[Fig.~\ref{fig3}(b)] show that all states are mostly localized at the edges of
the dot and are sublattice polarized as well.
Here, because atoms of the dot edges belong to the B-sublattice, carriers are 
only localized at the B atoms [open circles in Fig.~\ref{fig3}(b)].
Furthermore, the probability densities of each energy cluster are almost evenly 
distributed between the two layers.

Including the MF Hubbard model, one can see that each (spin-degenerate) 
energy state is now spin polarized.
Filled blue (empty red) symbols in Fig.~\ref{fig3}(a) show the spin up (down) energy levels.
As seen, each energy state in the two HOMOs and LUMOs clusters are considerably
affected by the e-e interaction.
However, the same type of SP energy degeneracy is still preserved for each 
spin-polarized energy levels.
The spin-polarized energy gap is $\Delta_{H} \approx 0.15$ eV.
Local magnetic moments $m_i$, shown in Fig.~\ref{fig3}(c), demonstrate that the
two layers are ferromagnetically coupled to each other, and each layer shares the same 
magnetization behavior.
However, the magnetic moments of the two sublattices in each layer are antiferromagnetic ordered.
Further, the moments corresponding to the edges' atoms are the largest, which decay to 
zero in the center of the dot.
The same behavior was also seen in both triangular MLG and AA-stacked BLG QDs
\cite{Fernandez2007,Nascimento2017}.
The total spin of the dot is $S = 7$, which agrees with Lieb's theorem, 
which states that a bipartite system described by the Hubbard model at half-filling
displays a ground state with a net spin of magnitude \cite{Lieb1989}
\begin{equation} \label{eq:Lieb}
 S = \frac{1}{2} |N_A - N_B|.
\end{equation}
This theorem allows us to predict the spin of the ground state of the bipartite
molecular systems, such as graphene-based nanostructures, only by counting the
sublattice imbalance $|N_A - N_B|$.

After investigating the energy states for a particular AA-like dot, now 
we examine the effect of twisting angle $\theta$ on the variation of the energy levels 
as a function of $\theta$ for fixed dot sizes.
Figure \ref{fig4} shows the lowest energy levels around $E_f = 0$ for different
dot sizes with the number of edge atoms $N_s = 2,3, \ldots,$ and $8$.
The results depict the two clusters of the HOMOs and LUMOs of the SP framework 
(black dashed) and the corresponding spin-polarized energy levels [spin up (blue) and 
spin down (red)] of the Hubbard model.
First, notice that the SP energy spectrum features an equal opening energy gap independent 
of the dot sizes when $\theta = 0$, which corresponds 
to the AA-stacked configuration of the triangular tBLG QD.
The size of the SP energy gap at $\theta = 0$ is $\Delta_{SP} = \epsilon^+ - \epsilon^- = 0.74$ eV, 
($\epsilon^\pm \approx \pm 0.37$ eV), which is the largest value for the entire range of 
twist angle.
This is expected since the untwisted structure (AA stacking) has the largest 
interlayer coupling between edge atoms, see Fig.~\ref{fig1}(c).
However, as seen in Fig.~\ref{fig4}, increasing the twisting angle leads the 
energy gap to shrink and close (or become minimum) at certain values of $\theta$. 
Notice that the gap decreases quickly and exhibits an oscillatory behavior for 
large dot sizes as the twist angle increases.
Further, the maximum energy separation between HOMOs and LUMOs at $\theta = 60^\circ$
occurs for $N_s^{(2)}$-group QDs, as seen for $N_s = 2,5$ and $8$ in 
Figs.~\ref{fig4}(a), \ref{fig4}(d), and \ref{fig4}(g), respectively.
This can be understood by the edge atoms' coupling, whose $N_s^{(2)}$-group dots are
greater than those of the two others.
In this case, as seen in Fig.~\ref{fig2}(a), the edge atoms of each layer are directly 
connected to the atoms of the adjacent layer that belong to the sublattice of the edge atoms,
i.e. B1-B2.
As previously shown, the HOMOs and LUMOs probability densities are sublattice polarized 
and solely localized at the B1 and B2 sublattices, which leads to a strong coupling 
between the layers in $N_s^{(2)}$-group dots.

The corresponding total spin $S$ for the dots with the spin-polarized energy levels 
shown in Fig.~\ref{fig4}, is plotted in Fig.~\ref{fig5}(a).
All dot sizes for small twist angles (which decrease as dot size increases) have a total spin 
of $S = 0$ and the magnitude of $S$ is consistent with Lieb's theorem [Eq.~\eqref{eq:Lieb}] for 
large twisting angles.
The observed behavior for small twisting angles does not contradict the 
prediction of Lieb's theorem, but rather indicates that the energy levels 
at these angles are antiferromagnetically polarized, as can be seen 
directly from the energy spectra shown in Fig.~\ref{fig4}.
This antiferromagnetic spin alignment can be highlighted further by 
plotting the local magnetic moments for an example of dot size and 
twist angle, e.g., $N_s = 5$ and $\theta = 5.5^\circ$ [Fig.~\ref{fig5}(b)]. 
As seen, the spin polarization of each layer is oriented equally in opposite directions.
Accordingly, all dot sizes exhibit a critical value of twist angle ($\theta_c$), 
at which a magnetic phase transition occurs.
As visible in Fig.~\ref{fig5}(a), $\theta_c$ decreases as the dot size increases.

\begin{figure}
	\centering
	\includegraphics[width = 8.5 cm]{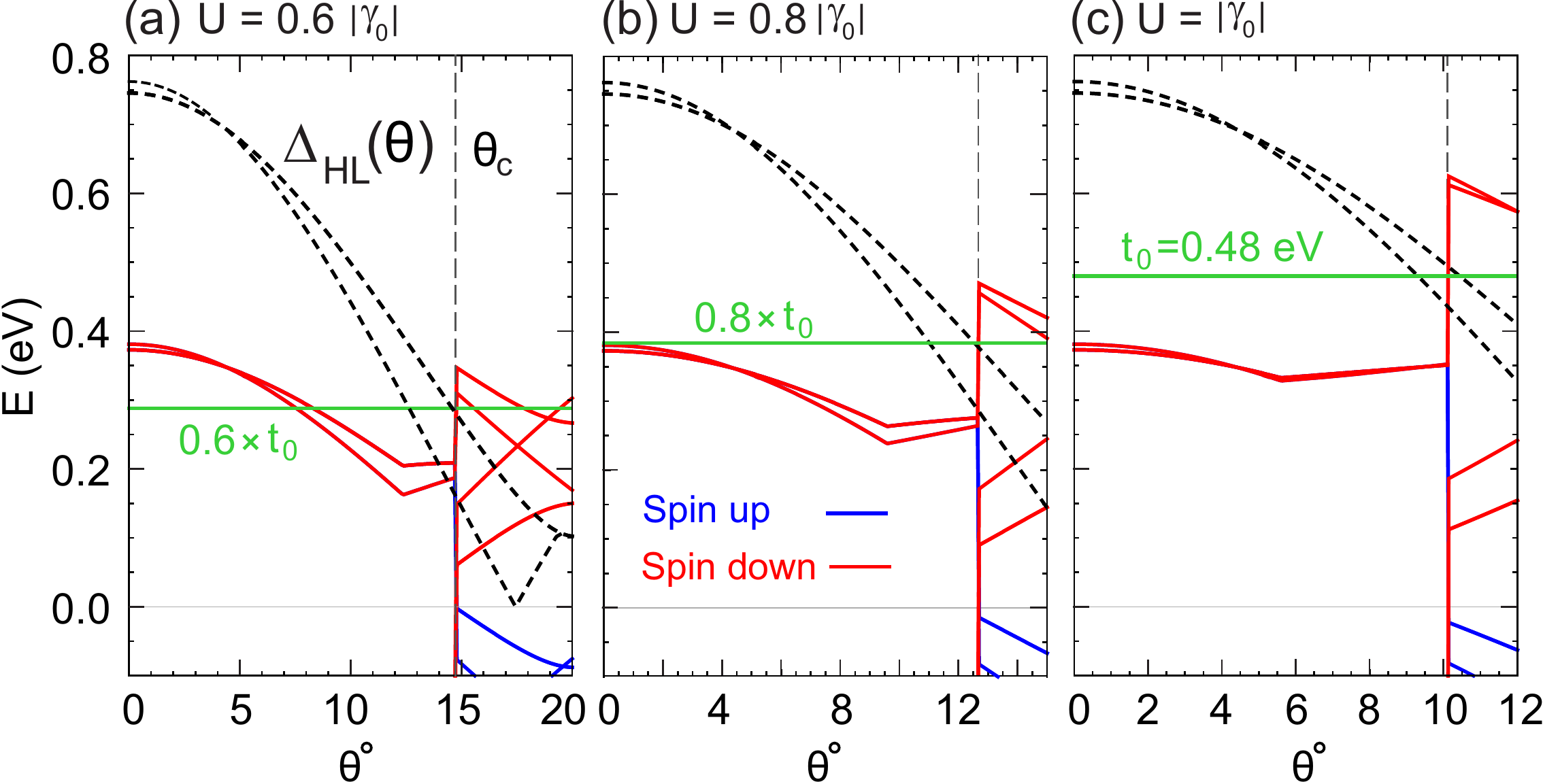}
	\caption{A zoom of the energy levels for an AA-like tBLG dot with $N_s = 4$, 
		as a function of $\theta$ for three different values of the Hubbard 
		parameter ($U = q\, |\gamma_0|$) (a) $q = 0.6$,	(b) $q =  0.8$, and (c) $q = 1$.
		As seen, when the maximum energy difference between two clusters of HOMOs and LUMOs
		($\Delta_{HL}(\theta)$; dashed curves) goes below $q \times t_0$ (green lines),
		the energy levels become spin polarized.
		}
	\label{fig6}
\end{figure}

\begin{figure}
	\centering
	\includegraphics[width = 8 cm]{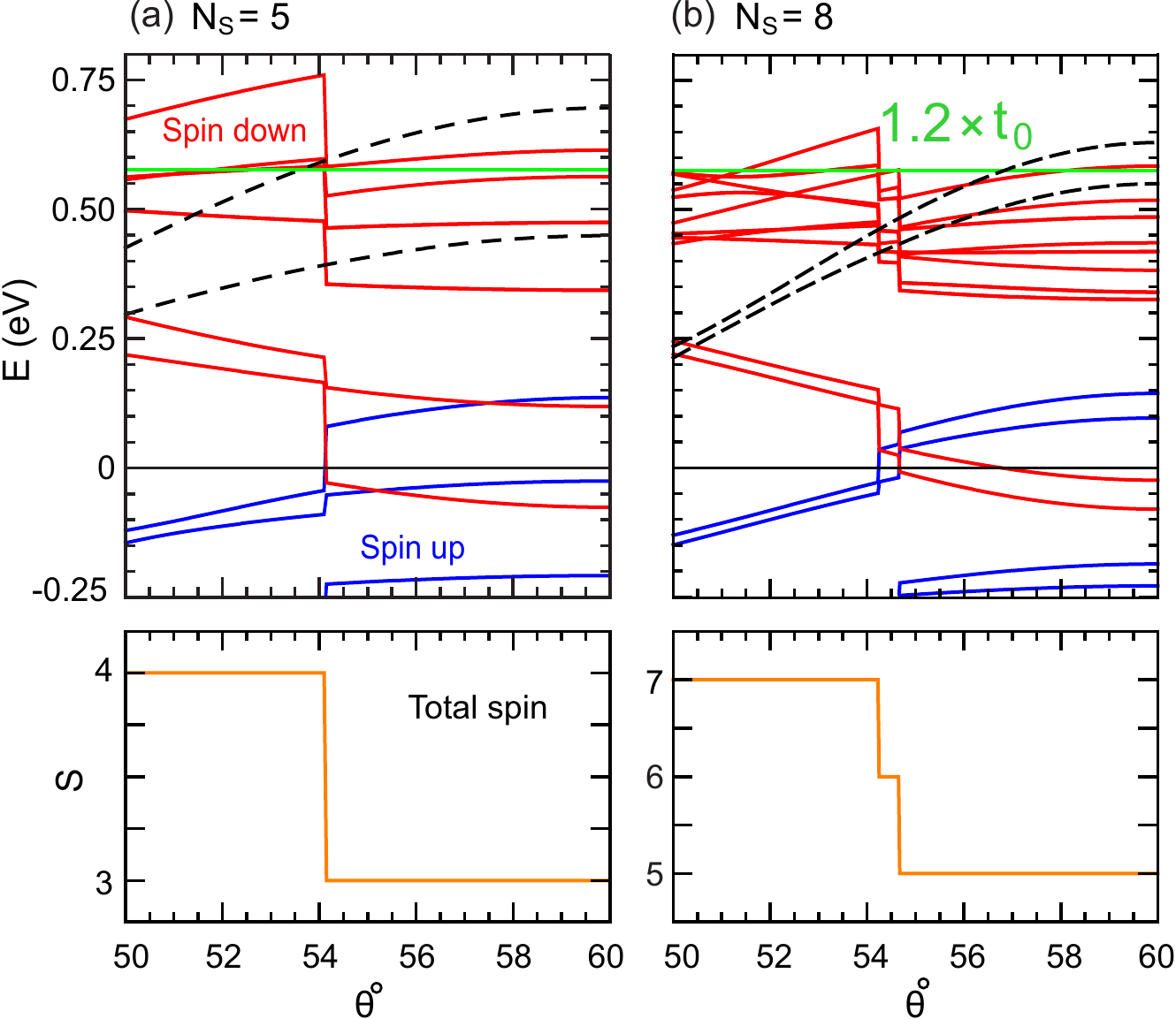}
	\caption{(Upper panel) A zoom of the lowest spin-polarized energy levels 
		of AA-like dots with (a) $N_s = 5$ and (b) $N_s = 8$ for the 
		twist angle range of $[50^\circ,~60^\circ]$.
		Dashed black curves show the energy gap of the two outmost SP HOMOs and LUMOs, 
		and the horizontal solid green line indicates $q \times t_0$ ($q = 1.2$).
		(Lower panel) The corresponding total spin $S$ for each dot.†
		}
	\label{fig7}
\end{figure}

Our numerical calculations demonstrate that the phase transition occurs when the ratio of 
the (maximum) energy difference between the HOMOs and LUMOs (in the SP frame), 
$\Delta_{HL}(\theta)$, to the \textit{interlayer} coupling $t_0$ becomes lesser than the 
ratio of the Hubbard parameter $U$ to the \textit{intralayer} hopping $|\gamma_0|$, i.e.,
\begin{equation} \label{eq:phase}
  \Delta_{HL}(\theta) \lessapprox q \times t_0,
\end{equation}
where $q = U / |\gamma_0|$.
To highlight this point, we plot, in Fig.~\ref{fig6}, a zoom of the Hubbard spin-polarized
energy levels and $\Delta_{HL}(\theta)$ for a dot with $N_s = 4$ and different 
Hubbard parameter values ($U = q\, |\gamma_0|$), e.g., (a) $q = 0.6$, (b) $q = 0.8$, 
and (c) $q = 1$.  
As seen, the above-mentioned criterion is met in all cases.
This behavior is also visible in the energy levels depicted in Fig.~\ref{fig4} 
for different dot sizes but with the constant $U = 1.2\, \gamma_0$ value.
Magnetic phase transition occurs for all dots when $\Delta_{HL} (\theta) / 2$ falls below 
$(q \times t_0) / 2 \approx 0.29$ eV (here $q = 1.2$); notice the horizontal green lines 
in all panels of Fig.~\ref{fig4} and the caption therein.

\begin{figure*}
	\centering
	\includegraphics[width = 18 cm]{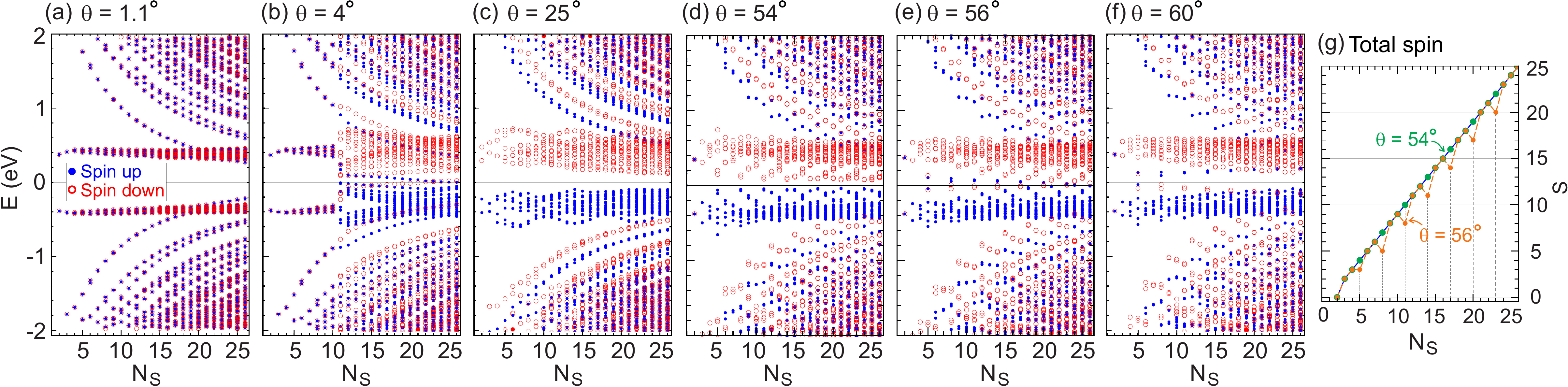}
	\caption{(a-f) Spin-polarized energy levels of the AA-like dot as a function of 
	dot size, characterized by $N_s$, around the Fermi energy $E_f = 0$  
	for six different twist angles as indicated in each panel. 
	Filled blue (empty red) circles represent spin-up (spin-down) energy levels.†
	(g) Net spin plotted for two twist angles of $\theta = 54^\circ$ (green) and 
	$\theta = 56^\circ$ (orange).
	The vertical dashed gray lines indicate $N_s^{(2)}$-group dots, whose net spin deviates 
	from Lieb's theorem prediction beyond the twist angle of $\theta \simeq 55^\circ$.
	}
	\label{fig8}
\end{figure*}

The spin-polarized energy levels exhibit smooth variation as a function of $\theta$ in the 
ferromagnetic phase for which the total spin $S$ 
of the structures agrees with the prediction of Lieb's theorem, as shown in Fig.~\ref{fig5}(a).
Notice that the net spin scales linearly with dot size by one spin unit.
While the $N_s^{(3),(4)}$-group dots exhibit a smooth variation of the energy spectra 
as the twist angle approaches $\theta = 60^\circ$, the energy levels of the 
$N_s^{(2)}$-group dots, i.e., $N_s = 2,5,$ and $8$, undergo an abrupt
decline in this area as seen in Figs.~\ref{fig4}(a), \ref{fig4}(d) and \ref{fig4}(g),
respectively.
Such abrupt drops in the energy levels are manifested as a decrease in the total spin of 
the dot by one or two units depending on the dot sizes, as shown in Fig.~\ref{fig5}(a).†
Figure \ref{fig7} shows a zoomed of the energy levels for two examples of these 
types of dots, i.e., $N_s = 5$ (a) and $N_s = 8$ (b), at twist angles between 
$50^\circ \le \theta \le 60^\circ$.
As seen, the abrupt decrease in energy levels around the twist angle of 
[$54^\circ$- $56^\circ$] results in an antiferromagnetically polarization of the lowest 
energy level(s).
This is a generic feature for $N_s^{(2)}$-group dots which will be discussed latter in 
Figs.~\ref{fig8}(d) and \ref{fig8}(e).
We attribute this behavior to the criterion mentioned in Eq.~\eqref{eq:phase}.
As depicted in Fig.~\ref{fig7}, once $\Delta_{HL}(\theta)$
(dashed black curves shown only for the two outermost HOMOs and LUMOs) exceeds $q \times t_0$ 
($\approx 0.58$ eV, green line), a decline in the energy levels occur.
This criterion is well matched in the case of $N_s = 5$ [Fig.~\ref{fig7}(a)], but there 
is some discrepancy for $N_s = 8$ [Fig.~\ref{fig7}(b)].

\begin{figure}
	\centering
	\includegraphics[width = 8.5 cm]{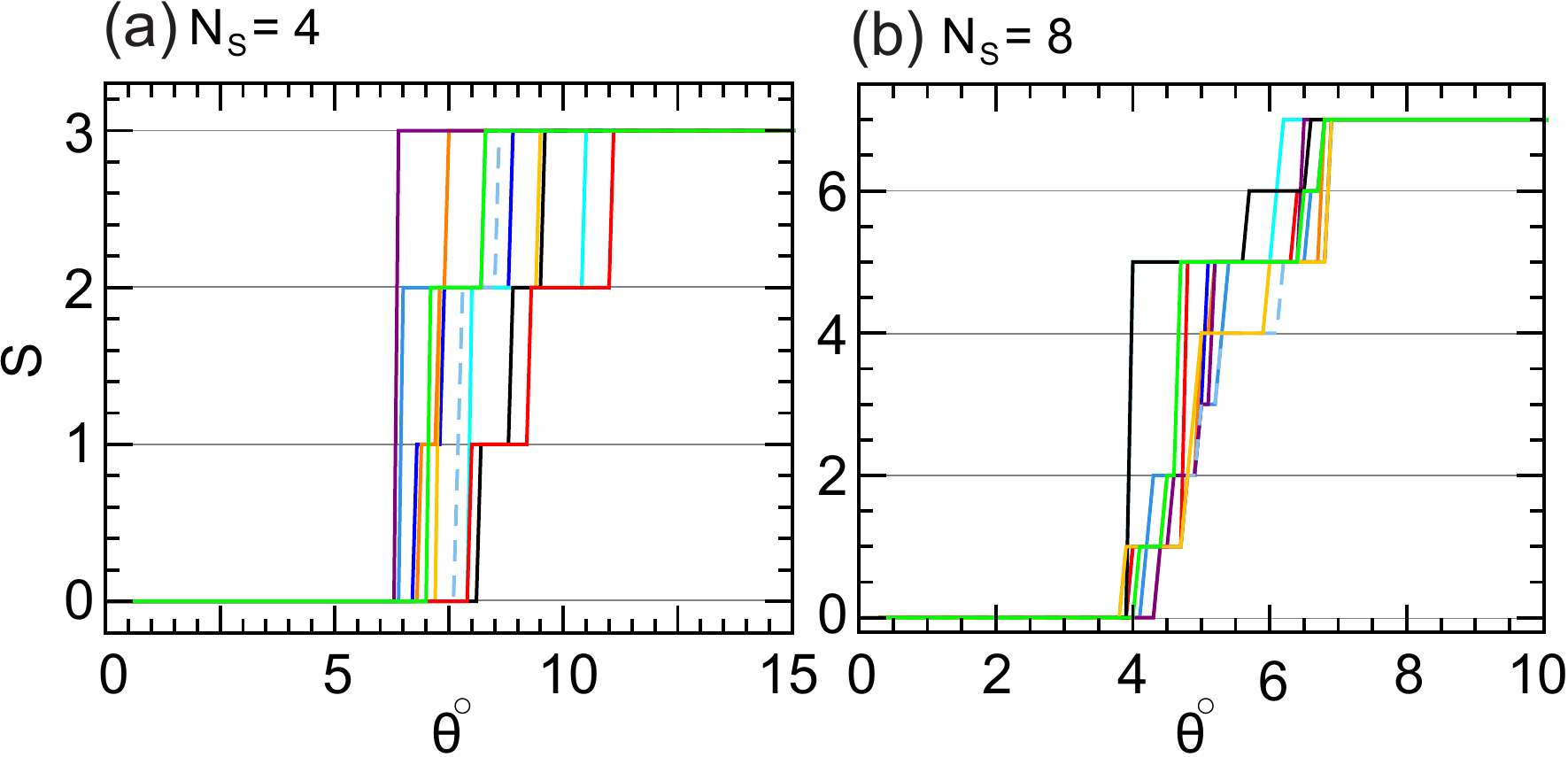}
	\caption{Total spin as a function of $\theta$ shown for ten different randomly chosen 
		initial electron densities $\langle n_{i \s} \rangle$ for two AA-like dot sizes 
		(a) $N_s = 4$ and (b) $N_s = 8$. †
	}
	\label{fig9}
\end{figure}

It is also interesting to discuss the dependence of magnetization on the dot size.
Figures \ref{fig8}(a)-\ref{fig8}(f) show the lowest energy levels as a function of the dot
size (characterized by $N_s$) for the six different twist angles 
$\theta = 1.1^\circ,\ 4^\circ,\ 25^\circ,\ 54^\circ, \ 56^\circ$, and $60^\circ$.
The results are shown up to $N_s = 26$.
As seen, for tiny twist angles, e.g., $\theta = 1.1^\circ$, the energy levels show 
antiferromagnetic polarization for the entire range of dot sizes.
In the case of $\theta = 4^\circ$ [Figs.~\ref{fig8}(b)],
small dots exhibit antiferromagnetic phase and turn to ferromagnetic phase when dot 
size increases, around $N_s = 11$.
This is because of the fast decline in the SP energy gap for the large dot sizes.
At intermediate twist angles, such as $\theta = 25^\circ$ [Fig.~\ref{fig8}(c)], 
the energy levels are perfectly aligned ferromagnetically; two clusters of spin-polarized
HOMOs and LUMOs are formed around the Fermi energy $E_f = 0$, where the energy gap
diminishes smoothly as the dot size increases.
This ferromagnetic phase is maintained for all dot sizes until $\theta \approx 55^\circ$, 
as seen in Fig.~\ref{fig8}(d).
However, beyond the angle $\theta \approx 55^\circ$ [Figs.~\ref{fig8}(e) and \ref{fig8}(f)], 
the $N_s^{(2)}$-group dots behave 
differently from the other two groups, with the lowest one or two energy levels being 
antiferromagnetically polarized, resulting in a drop in the net spin of the dots.
To highlight this further, total spin as a function of $N_s$ is shown in Fig.~\ref{fig8}(g)
for two twist angles $\theta = 54^\circ$ (green) and $\theta = 56^\circ$ (orange).
As seen, while the net spin of all dots scales linearly at $\theta = 54^\circ$, the 
$N_s^{(2)}$-group dots (marked by dashed gray vertical lines) 
show a reduced net-spin value from Lieb's theorem prediction for $\theta = 56^\circ$.
Except for $N_s = 2$, which displays one unit reduction, the remaining $N_s^{(2)}$-group dots
show a net spin of two units less than what Lieb's theory predicts.

\begin{figure*}
	\centering
	\includegraphics[width = 16 cm]{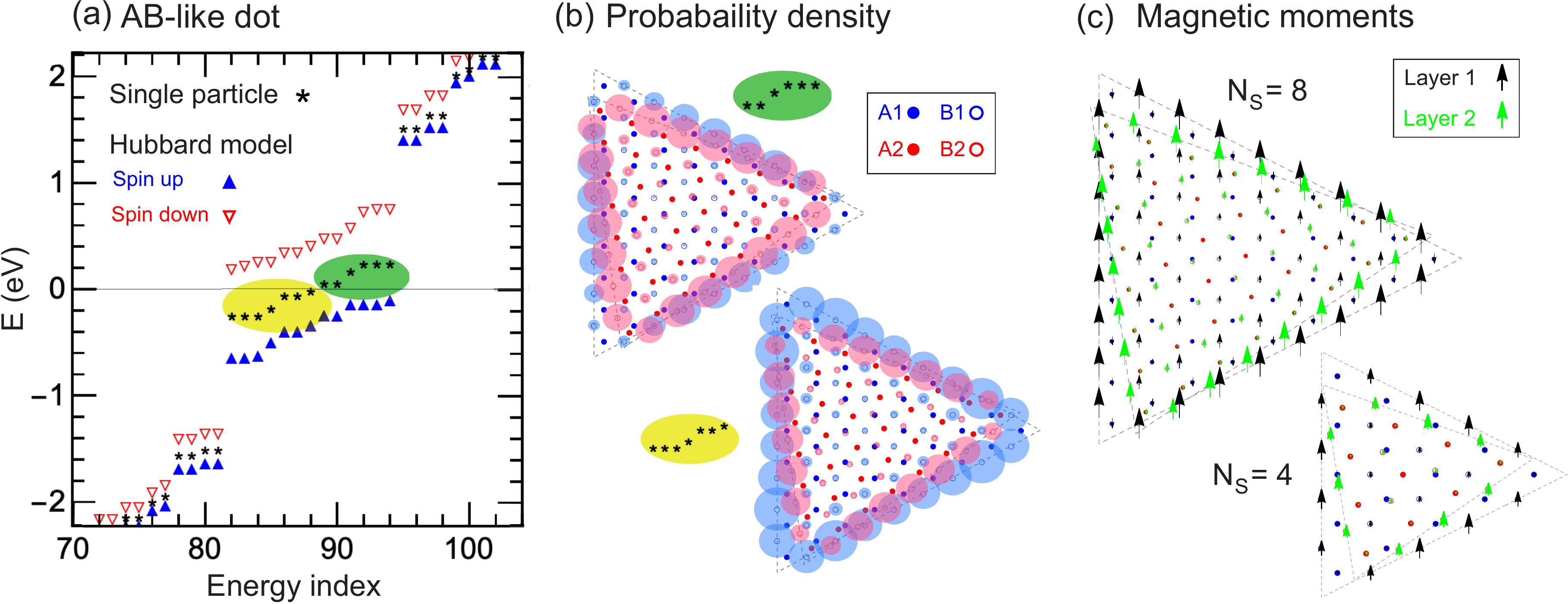}
	\caption{(a) Energy spectra of an AB-like dot with $N_s = 8$ and a twist 
		angle of $\theta = 7^\circ$ in the frame of SP (black stars) and Hubbard models 
		around the Fermi energy $E_f = 0$.
		Filled blue (empty red) symbols depict the up (down) energy levels.
		(b) Probability densities corresponding to two clusters of the LUMOs 
		(upper panel) and HOMOs (lower panel).
		Bottom (top) layer is represented by the light blue (red) color.  
		(c) Local magnetic moments $m_i$ for two examples of AB-like dots of $N_s = 8$
		and $N_s = 4$ ($\theta = 7^\circ$) are shown by the black and green arrows
		corresponding to the lower and upper layers, respectively.
		The length of the arrows is measure of the relative magnitude of the magnetic
		moments.†
	}
	\label{fig10}
\end{figure*}

At this point it is worth mentioning that all MLG carbon nanostructures studied
in previous literature feature zero-energy states for which any repulsive Coulomb interaction 
can cause spin-polarization, a mechanism for escaping an instability caused by the presence 
of low-energy electrons in the system \cite{Yazyev2010,Oteyza2022}.
As illustrated above, an AA-like dot no longer features strictly zero-energy states at 
small twist angles.
However, the same scenario occurs with such graphene QDs.
For small twist angles near the AA-stacking configuration, it is energetically 
favorable for spins in adjacent layers to couple to each other antiferromagnetically.
Increasing the twisting angle, HOMO and LUMO electronic states approach the Fermi level; 
in the case when the Coulomb repulsion and energy separation between HOMOs and LUMOs meet the 
criterion expressed by Eq.~\eqref{eq:phase}, the energy levels 
become spin polarized to reduce the density of states around the Fermi energy.
Furthermore, notice that, where the HOMOs and LUMOs  
are closer to the Fermi level at $E_f = 0$, spin-polarized energy levels are formed 
farther away, and vice versa, see Fig.~\ref{fig4}.

We also would like to point out an issue in the numerical calculation using the MF Hubbard
model for the studied dots.
Our findings show that the results for the critical values of twist angles 
at which the magnetic phase transition occurs can be varied depending on the randomly chosen
initial values for the electron densities $\langle n_{i \s} \rangle$.
For example, in Fig.~\ref{fig9}, we plot the net spin for two examples of the AA-like dots 
with (a) $N_s = 4$ and  (b) $N_s = 8$ for ten different randomly chosen electron densities 
$\langle n_{i \s} \rangle$.
As seen, the critical twist angles change slightly for different initial values of 
$\langle n_{i \s} \rangle$, and these variations appear to diminish as the dot size increases,
cf.\ Figs~\ref{fig9}(a) and \ref{fig9}(b).
All cases, however, share a similar thread.


\subsection{AB-like triangular tBLG QDs}

Now we investigate the magnetic properties of AB-like tBLG QD, whose configuration is
illustrated in Fig.~\ref{fig1}(b).
In this configuration, the top layer is smaller than the bottom layer by one edge atom, 
and $\theta = 0^\circ$ corresponds to the AB-stacking BLG QD [Fig.~\ref{fig1}(d)].
In BLG, the AB stacking configuration is more natural and stable than the AA arrangement.
As we discuss below, such a BLG dot features an odd number of degenerate edge
states that can generate a half-integer net spin in contrast to the AA-like dots with 
an integer net spin.

Figure \ref{fig10}(a) shows the energy levels for an example of AB-like tBLG QDs with 
the same parameters as used for the AA-like dot of previous section [Fig.~\ref{fig3}(a)], 
i.e., $N_s = 8$ and $\theta = 7^\circ$.
Here also, the number of SP edge states is consistent with the benzenoid graph theory.
As seen, there are $13$ edge states (black stars): 
$N_Z = 7$ (encircled in the yellow oval) from the bottom layer and $N_Z = 6$ 
(green oval) from the upper layer.
Notice that here the edge states are more dispersed compared to those in the AA-like 
configuration [cf.\ Figs.~\ref{fig3}(a) and \ref{fig10}(a)].
The SP energy gap is $\Delta_{SP} \approx 0.08$ eV, which is smaller than for the AA-like 
dot by a factor of $3.5$.
This is related to the coupling between edge atoms, which is weaker in the AB-like dots.†
The probability densities corresponding to each layer's edge states are 
mostly concentrated in the same layer [Fig.~\ref{fig10}(b)], as opposed to the 
AA-like dot, whose edge states are almost evenly distributed in both layers 
due to layer symmetry [Fig.~\ref{fig3}(b)].
However, the edge-state densities in both types of dots are sublattice polarized.

Similar to the case of the AA-like dot, including MF on the level of the Hubbard model, 
the edge states are considerably affected by the e-e interaction. 
Here, the spin-polarized energy gap is about $\Delta_{H} \approx 0.60 $ eV, which is four times
larger than that in the AA-like dot.
The local magnetic moments are depicted in Fig.~\ref{fig10}(c), demonstrating that 
in the bottom layer they are somewhat larger than that in the top layer.
This feature is more pronounced for small dot sizes as shown in Fig.~\ref{fig10}(d)
for a dot with $N_s = 4$.

\begin{figure}
	\centering
	\includegraphics[width = 8.5 cm]{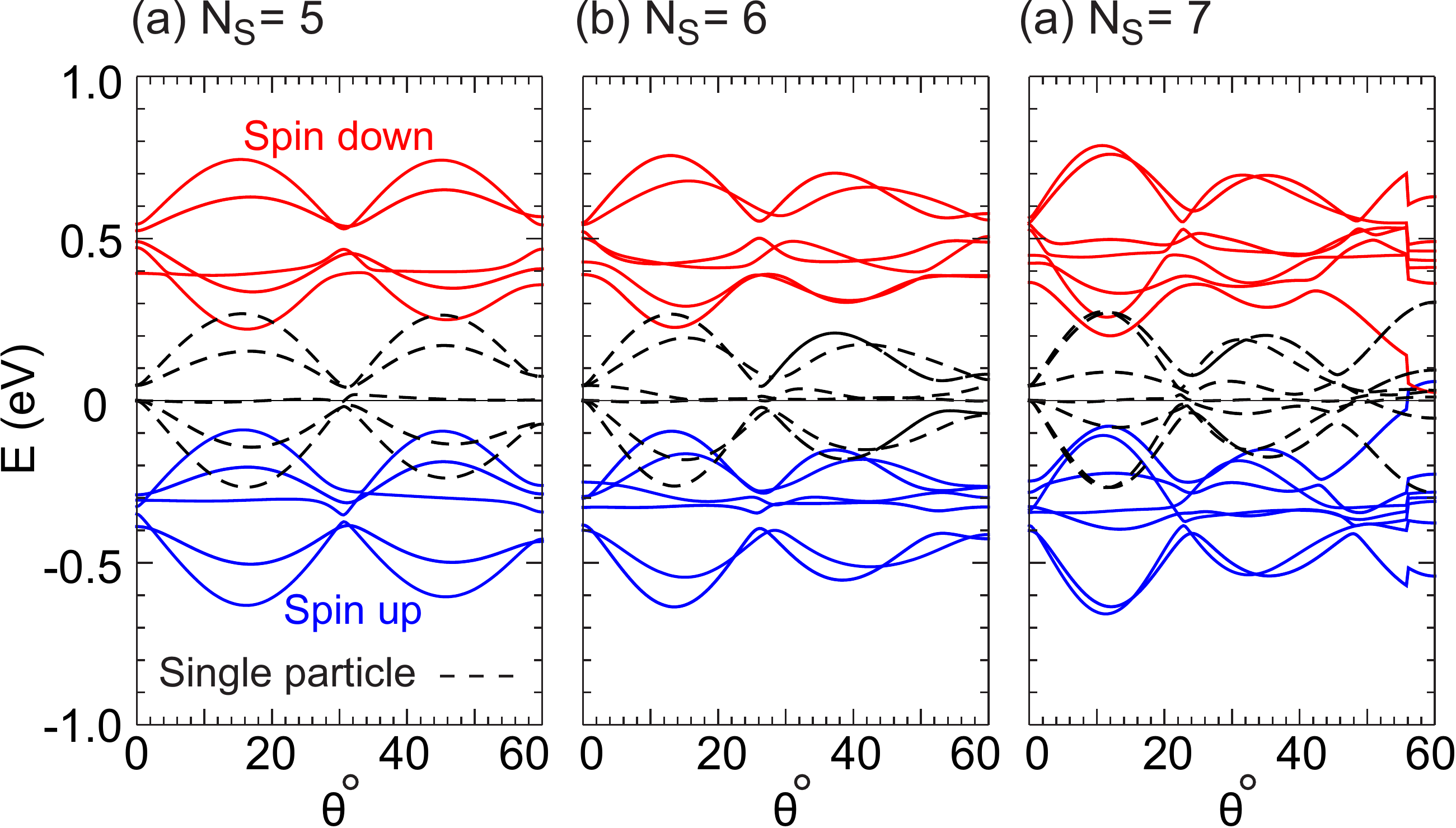}
	\caption{Energy levels of the AB-like dots as a function of twist angle 
		$\theta$ for three different dot sizes as indicated by the number $N_s$ of edge atoms  
		in each panel. 
		Dashed black curves are for SP model, and blue and red ones show the spin up 
		and spin down energy levels in the Hubbard model.†
	}
	\label{fig11}
\end{figure}

In Fig.~\ref{fig11}, we show the the lowest energy levels around $E_f = 0$ 
for three dot sizes of the AB-like dots with the edge atoms of (a) $N_s = 5$, 
(b) $N_s = 6$, and (c) $N_s = 7$, respectively, representing 
$N_s^{(2), (3), (4)}$-group dots [Eq.~\eqref{eq:categ}].
As seen, all dots exhibit ferromagnetically-polarized energy levels throughout the whole 
twist angle range of $[0, 60^\circ]$, with the net spin $S$ consistent with 
Lieb's theorem prediction.
The numerically calculated total spins are shown in each panel.
Notice that because of the one edge atom discrepancy between the bottom and top layers in the 
AB-like dots, the total spins come out as half-integer numbers.
However, here too, the $N_s = 7$ dot (an example of $N_s^{(4)}$-group dots) 
undergo abrupt drops in the energy levels when the twist angle approaches $\theta = 60^\circ$.
This behavior is analogous to that of the $N_s^{(2)}$ group of the AA-like dots, as 
explained in Fig.~\ref{fig7}.
This is to be expected as we can see from the geometries in Figs.~\ref{fig2}(a) and 
\ref{fig2}(f), which are both similar around $\theta = 60^\circ$.

Finally, we conclude this section by mentioning a few words about the magnetism in MLG and BLG
nanostructures.
In contrast to MLG triangle QDs \cite{Yazyev2010}, where the total spin scales linearly with 
dot size by half a 
spin unit, i.e., $S = 1/2, 1, 3/2, \ldots$, the examples above show that the $S$ for both tBLG QD 
configurations in the ferromagnetic phase scales linearly with dot size by one spin unit.
However, as discussed above, AA-like and AB-like dots feature integer
($S = 1, 2, 3, \ldots$) and half-integer ($S = 3/2, 5/2, \ldots$) total spin, respectively.
A striking feature of a tBLG QD is its capacity to control the position of its energy levels
by tuning the relative twist angle of the layers while still representing a specific spin.
Besides, the prediction of a magnetic quantum phase transition at a critical twisting angle
would be of interest for understanding the fundamental physics of magnetism in tBLG dots, 
as well as for potential applications in areas such as spintronics and quantum computing.

\section{Conclusion} \label{con}
In conclusion, using the TBM in combination with the MF Hubbard model, we studied the magnetic
properties of zigzag-edged triangular QDs in tBLG with a focus on the 
effects of variations in the twist angle as well as dot size.
We considered two configurations of tBLG QDs: AA-like and AB-like QDs, whose untwisted 
arrangements correspond, respectively, to the AA- and AB-stacked BLG QDs.
Depending on the dot size, such QDs would have a different geometries, and we classified 
them appropriately into three categories.

%
Our findings show that the AA-like dots exhibit an antiferromagnetic phase for small twist 
angles, which transits to a ferromagnetic phase beyond a critical value of $\theta_c$.†
Our analysis shows that the size of $\theta_c$ decreases as the dot size increases.
We also found a criteria for such $\theta_c$, according to which the dots exhibit 
ferromagnetic spin polarization as long as the energy difference between the electron 
and hole edge states (in the single particle frame) is less than 
$(U / \gamma_0) \times t_0$, where $U$ 
is the Hubbard e-e interaction and $\gamma_0$ ($t_0$) denotes the graphene 
\textit{intralayer} (\textit{interlayer}) hopping.
Unlike AA-like dots, spins in the AB-like dots are ferromagnetically polarized for the 
entire range of twist angle.
The net spin $S$ of both types of  QDs in the ferromagnetic phase is consistent with the prediction
from Lieb's theorem, where AA-like (AB-like) dot exhibits an integer (half-integer) total spin value. 

Due to the dispersive and oscillatory behavior of the energy levels as a function of 
twist angle in such QDs, the ferromagnetic phase is preserved as long as the energy gap 
between the edge states is satisfied by the above-mentioned criterion.
Our analysis showed that depending on whether the entire or part of such gaps surpass the 
amount of $(U / \gamma_0) \times t_0$, the spins can be polarized antiferromagnetically with 
$S = 0$ or ferromagnetically with a finite $S$ (an integer multiple of 1/2) less than Lieb's 
theorem prediction.
For the applied Hubbard e-e interaction here ($U = 1/2\, \gamma_0$), the latter property 
occurred for a group of both types of dots at the twist angle range of 
$[55^\circ-60^\circ]$, i.e., when two triangular layers are almost overlapped in the opposite 
direction, as in the \dql David star\dqr\ configuration.

Using the twist angle as a knob to tune the energy levels of QDs in tBLG presents 
an interesting opportunity to manipulate the charge and spins in such nanostructures, which
are promising candidates for future electronic and spintronic technologies.

\section*{Acknowledgments}

This work was supported by the Institute for Basic Science in Korea (No. IBS-R024-D1).
D.R.C. is grateful to the National Council of Scientific and Technological Development 
(CNPq, supported by grand number 313211/2021-3) and the National Council for the Improvement
of Higher Education Personnel (CAPES) of Brazil for financial support.


\end{document}